\documentclass[conference]{IEEEtran}
\IEEEoverridecommandlockouts
\usepackage{cite}
\usepackage{amsmath, amssymb, amsfonts, subcaption, bm}
\usepackage{graphicx}
\usepackage{textcomp}
\usepackage{xcolor}

\def\BibTeX{{\rm B\kern-.05em{\sc i\kern-.025em b}\kern-.08em
    T\kern-.1667em\lower.7ex\hbox{E}\kern-.125emX}}
\begin{document}

\title{Log-FSK: A Frequency Modulation for Over-the-Air Computation
\thanks{This work is part of the project IRENE (PID2020-115323RB-C31), funded by MCIN/AEI/10.13039/501100011033 and supported by the Catalan government through the project SGR-Cat 2021-01207.}
}

\author{\IEEEauthorblockN{Marc Martinez-Gost\IEEEauthorrefmark{1}\IEEEauthorrefmark{2}, Ana Pérez-Neira\IEEEauthorrefmark{1}\IEEEauthorrefmark{2}\IEEEauthorrefmark{3}, Miguel Ángel Lagunas\IEEEauthorrefmark{2}}
\IEEEauthorblockA{
\IEEEauthorrefmark{1}Centre Tecnològic de Telecomunicacions de Catalunya, Spain\\
\IEEEauthorrefmark{2}Dept. of Signal Theory and Communications, Universitat Politècnica de Catalunya, Spain\\
\IEEEauthorrefmark{3}ICREA Acadèmia, Spain\\
\{mmartinez, aperez, malagunas\}@cttc.es}
}

\maketitle

\begin{abstract}
In this study we introduce Logarithmic Frequency Shift Keying (Log-FSK), a novel frequency modulation for over-the-air computation (AirComp). Log-FSK leverages non-linear signal processing to produce AirComp in the frequency domain, this is, the maximum frequency of the received signal corresponds to the sum of the individual transmitted frequencies.
The demodulation procedure relies on the inverse Discrete Cosine Transform (DCT) and the extraction of the maximum frequency component.
We provide the theoretical performance in terms of error probability and mean squared error.
To demonstrate its practicality, we present specific applications and experimental results showcasing the effectiveness of Log-FSK AirComp within linear Wireless Sensor Networks (WSN). Our experiments show that Log-FSK outperforms linear AirComp, implemented with double sideband (DSB), when working above the threshold SNR.
\end{abstract}


\section{Introduction}
Over-the-air computation (AirComp) has gained attention in the context of distributed function computation \cite{Nazer2007}.
The key behind AirComp is designing codes considering the communication and computation simultaneously. Specifically, AirComp exploits that node identity is irrelevant for many computations (e.g., statistics) and that the additive nature of the communication channel can be used for simultaneous transmission in time and frequency, without the need of orthogonal resource allocation.
In this respect, AirComp is suitable for next generation multiple access (NGMA) schemes, as it removes the orthogonality of resource blocks, provides high spectral efficiency as well as low latency.

Nonetheless, most of the literature in AirComp assumes a baseband signal model \cite{sahin22}, which ignores the need of a modulation that matches the additive nature of the communication channel. In other words, these works implicitly assume a linear modulation (e.g., double sideband) where the information is carried in the amplitude of the waveform. It is widely known that these modulations are highly susceptible to noise and interferences, have limited bandwidth efficiency and require more complex demodulation schemes than digital modulations.
However, the design of modulation for AirComp has been overlooked in the literature.

The applicability of existing modulations, which are mostly angular, to AirComp is challenging. With these frequency and phase modulations, the additive nature of the communication channel does not translate into the sum of angular information.
In this respect, in \cite{razavikia23} the authors propose to redesign digital modulations for AirComp. The signal constellation is constructed for each function value and the demodulation provides the desired aggregated output. However, constellations are not dense and its application to analog data is still to be explored.
In \cite{martinez23} the authors propose to use a type-based multiple access (TBMA) with frequency modulations for function estimation. However, this does not fully exploit AirComp because several communication resources (e.g., frequency carriers) are used.

In this work we propose Logarithmic Frequency Shift Keying (Log-FSK), a novel waveform for AirComp that relies on digital FSK to carry analog data. To overcome the concerns of linear analog modulations, Log-FSK carries information only in the frequency of the signal, not in the amplitude. The modulation is designed so that a non-linear processing at the receiver yields the sum of modulated information in the frequency, this is, the resulting frequency is the sum of individual transmitted frequencies.
We highlight the fact that, unlike linear AirComp, FSK modulations are currently used in communication networks. An instance is the Long Range (LoRa) modulation \cite{Chiani2019}. Besides, Log-FSK does not require changing the overall architecture, but append additional blocks to standard communication systems. 
We provide the theoretical performance of Log-FSK in terms of error probability and mean squared error (MSE), and evaluate its performance in a communication channel for different number of simultaneous transmitters.

The remaining part of the paper proceeds as follows:
Section II starts with a review of the literature of modulations for AirComp.
Section III presents the Log-FSK waveform and characterizes its spectrum.
Then, section IV analyses the performance of Log-FSK in an AWGN and fading channels. Finally, Section V proposes specific computation of functions and complements the theoretical analysis with experimental results. Section VI concludes the paper.


\section{Waveform Design}
\label{sec:waveform_design}
\begin{figure}[t]
\centering
\includegraphics[width=\columnwidth]{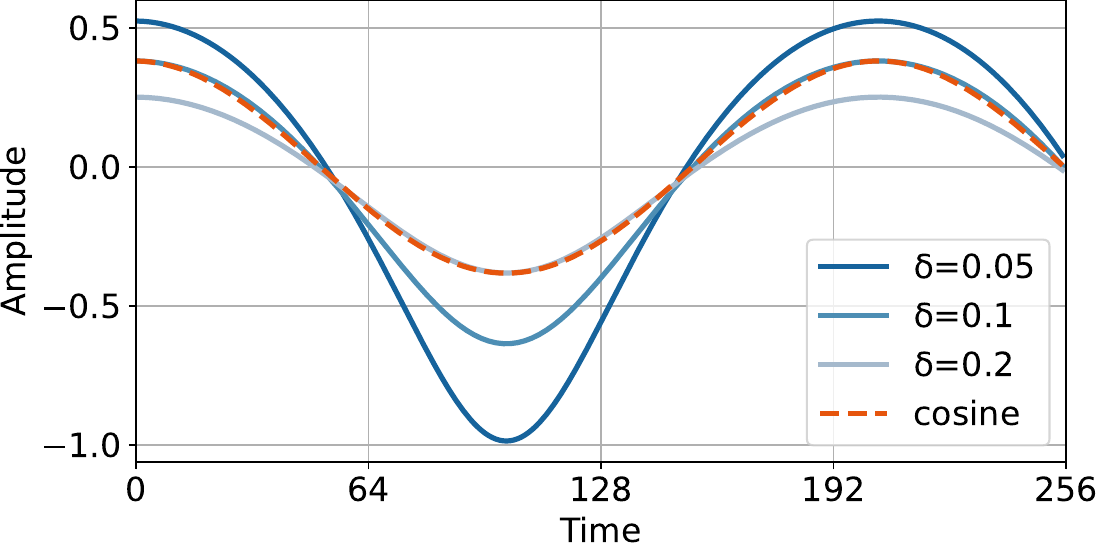}
\caption{Proposed logarithmic waveform for different $\alpha=\sqrt{2/N}+\delta$ and a cosine waveform at frequency $m=5$.}
\label{fig:logos_waveform}
\vspace{-0 pt}
\end{figure}

Consider the following discrete-time signal:
\begin{equation}
    x[n] = \log\left(\sqrt{\frac{2}{N}}\cos\left(\frac{\pi(2m+1)}{2N} n\right)
    +\alpha\right)
    \label{eq:logos_waveform}
\end{equation}
for $n=0,\dots,N-1$, where $m\in[0,N-1]\subset{\mathbb{N}}$ is the discrete frequency, $N$ is the number of samples and $\log(\cdot)$ is assumed to be the natural logarithm without loss of generality. Parameter $\alpha$ is a regularization factor that prevents the argument of the logarithm from reaching negative values, specifically $\alpha>\sqrt{2/N}$. 
The mean of $x[n]$ may be subtracted so that the signal has no DC component. Fig. \ref{fig:logos_waveform} shows the waveform along with the corresponding cosine at the same frequency ($m=5$, $N=256$). The proposed signal retains the same periodicity as the cosine due to the the monotonicity of the logarithm.

\subsection{Log-FSK modulation}
\label{sec:modulation}
Consider a set of $K$ users, where $m_k$ is the measurement of user $k$, which is modulated as in $M$-ary Frequency Shift Keying ($M$-FSK) with $M=N$, but using the waveform in \eqref{eq:logos_waveform}:
\begin{align}
    x_k[n] &= \log\left(\sqrt{\frac{2}{N}}\cos\left(\frac{\pi(2m_k+1)}{2N}n\right)+\alpha\right)\nonumber\\
    &=\log\left(\cos_{m_k}[n]+\alpha\right),
    \label{eq:logos_modulation}
\end{align}
in which we defined
\begin{equation}
    \cos_{m_k}[n]=
    \sqrt{\frac{2}{N}}\cos\left(\frac{\pi(2m_k+1)}{2N}n\right)
    \label{eq:cosine}
\end{equation}
The cosines correspond to the discrete cosine transform (DCT) basis (see \cite{Gost23}), and we refer to the modulation in \eqref{eq:logos_modulation} as the Logarithmic FSK (Log-FSK). Consider two users, modulating their respective measurements, $m_k$ and $m_l$, as in \eqref{eq:logos_modulation}. For simplicity, consider an ideal noiseless channel and both users transmitting concurrently and perfectly synchronized in time. Upon reception, the receiver computes the exponential over the received signal as
\begin{align}
    r[n] &=\exp{\left(x_k[n]+x_l[n]\right)}\nonumber\\
    &=
    \frac{1}{2}\sqrt{\frac{2}{N}}\left(\cos_{m_k+m_l}[n] + \cos_{|m_k-m_l|}[n]\right)\nonumber\\
    &\,\quad
    +\alpha\cos_{m_k}[n]+
    \alpha\cos_{m_l}[n] +\alpha^2
    \label{eq:FM_AirComp}
\end{align}

The first term in \eqref{eq:FM_AirComp} reveals that AirComp using frequency modulations is, at least, theoretically possible. In Sec. \ref{sec:Communication} we will generalize this result for noisy channels and an arbitrary number of transmitters. The key step for achieving AirComp using frequency modulations is to transform the additive channel into a multiplicative channel. Using the Kolmogorov superposition theorem \cite{Kolmogorov}, the multiplication of distributed signals can be achieved by a logarithmic preprocessing at each transmitter and an exponential postprocessing at the receiver. We have used this insight to integrate the logarithmic processing in the waveform design and develop \eqref{eq:logos_waveform}.


\subsection{Log-FSK demodulation}
\begin{figure}[t]
\centering
\includegraphics[width=\columnwidth]{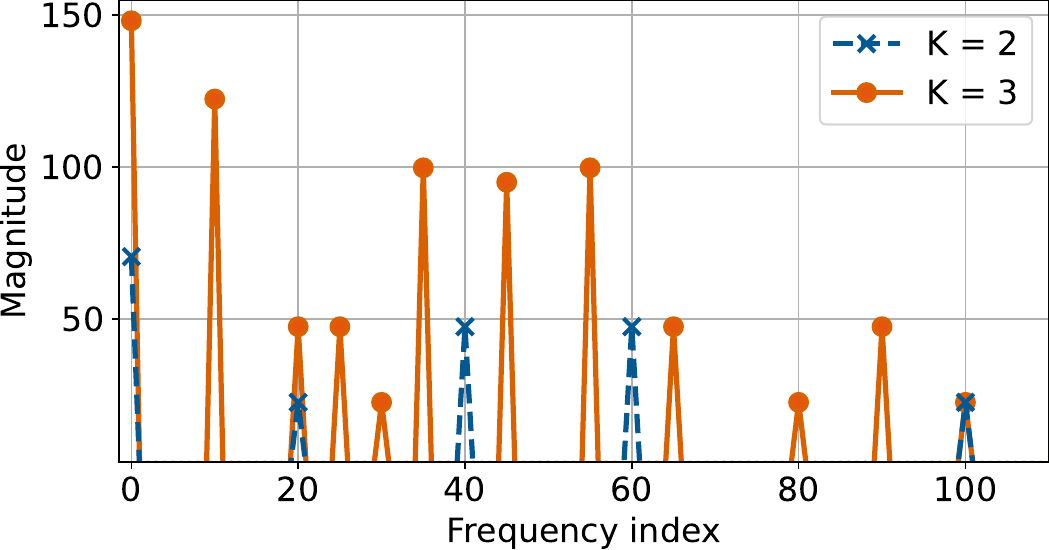}
\caption{Signal after demodulation for $K=\{2,3\}$. The modulated information is $[40, 60]$ and $[10, 35, 55]$, respectively. In both cases the maximum frequency is located at the sum of the transmitted information, this is, $100$.}
\label{fig:logos_demodulation}
\vspace{-0 pt}
\end{figure}

When the frequency of the sum in \eqref{eq:FM_AirComp} is guaranteed to belong to the original domain, i.e., $\Sigma\triangleq m_k+m_l\in[0,N-1]$, it is easy to see that $\Sigma$ corresponds to the maximum frequency of \eqref{eq:FM_AirComp}. The extension to $K$ users and under the same constraint is straightforward. 
The previous insight allows to design the demodulation process, since it requires detecting the maximum frequency. Provided that the modulation \eqref{eq:logos_modulation} contains the DCT basis, we will uses the inverse DCT over $r[n]$ (i.e., a bank of filters) to recover its spectral components.
Fig. \ref{fig:logos_demodulation} shows the inverse DCT applied over $r[n]$ for $K=\{2,3\}$ transmitters. The modulated measurements are $\mathbf{m}=[40,60]$ and $\mathbf{m}=[10,35,55]$, respectively. Besides all the intermodulation products, in both cases the maximum frequency is located at the sum, namely,
$\Sigma=\sum_{k=1}^{K}m_k=100$.
In the presence of noise, this detection problem requires thresholding $r[n]$ to remove the noise. In the following we will characterize the noise and define an appropriate thresholding.


\section{Log-FSK in a Communication Channel}
\label{sec:Communication}
In this section we will analyze the performance of Log-FSK in an additive white Gaussian noise (AWGN) channel.
Due to the non-linear processing required at demodulation, Log-FSK demands tight power control. Otherwise, any slight difference in the transmitted power would introduce non-linear undesired effects in \eqref{eq:FM_AirComp}. For this reason, any channel fading needs to be compensated as in analog AirComp.

\subsection{AWGN channel}
For the sake of notation simplicity and without loss of generality we constrain the analysis to $K=2$ transmitters. Consider that both users control the transmitted power using amplitudes $A_c$ and $B_c$ as
\begin{align}
    x_k[n] =A_c\log\left(B_c\cos_{m_k}[n]+\alpha\right)
    \label{eq:logos_modulation_amplitude}
\end{align}

We further define $A_c=\bar{A}_c/\sqrt{P}$, with $P$ being the power of the logarithmic term. In this respect, $\bar{A}_c$ is independent of $B_c$ and the transmitted power equals $\bar{A}_c^2$. Notice that $\alpha$ now depends on $B_c$ and must satisfy $\alpha>B_c\sqrt{N/2}$.

The signal at the input of the receiver is
\begin{equation}
    y[n] = x_k[n] + x_l[n] + w[n],
    \label{eq:signal_rx}
\end{equation}
where $w$ is distributed as $\mathcal{N}(0,\sigma^2)$ and $\sigma^2$ is the noise power. We omit standard communication blocks (i.e., downconversion, sampling, etc.) Before applying the exponential postprocessing the receiver needs to compensate for $A_c$, resulting in
\begin{align}
    r[n] &=\exp{\left(y[n]/A_c\right)}\nonumber\\
    &=
    \Biggl[\frac{B_c^2}{2}\sqrt{\frac{2}{N}}\left(\cos_{m_k+m_l}[n] + \cos_{|m_k-m_l|}[n]\right)\nonumber\\
    &\quad\quad\quad\quad
    +\alpha B_c\left(\cos_{m_k}[n]+
    \cos_{m_l}[n]\right) +\alpha^2\Biggr]z[n],
    \label{eq:FM_AirComp_noise}
\end{align}
where $z[n]=\exp{(w[n]/A_c)}$ follows a log-normal distribution, $z[n]\sim\log\mathcal{N}(0,\sigma^2/A_c^2)$ with mean and variance
\begin{align}
    \mu_z&=\exp(\sigma^2/2A_c^2)\\
    \sigma^2_z&=\exp(\sigma^2/A_c^2)\left(\exp(\sigma^2/A_c^2)-1\right)
\end{align}
The multiplicative channel designed for frequency-based AirComp does not come for free, as the noise is now multiplicative and not normally distributed. A standard technique to deal with multiplicative noise is homomorphic filtering, in which the signal is transformed into a domain where the noise becomes additive and a linear filter can be applied \cite{Opp04}. This corresponds to applying a logarithmic transformation over $r[n]$, which converts the product into a sum; however, this procedure cannot be followed in this scenario because Log-FSK requires a multiplicative channel. Thus, Log-FSK has to deal with multiplicative noise, which will degrade the system performance unless high signal-to-noise ratio (SNR) is guaranteed.

\subsubsection{Received \textup{SNR} $\textup{(SNR}_R$\textup{)}}
The SNR at the input of the receiver (i.e., predetection SNR) and corresponding to a single transmitter is $\text{SNR}_R= \bar{A}_c^2/\sigma^2$. While the total power increases linearly with the number of users, we define the SNR with respect to each transmitter for comparison purposes.

\subsubsection{Destination \textup{SNR} $\textup{(SNR}_\Sigma$\textup{)}}
In order to find the SNR at the destination, we particularly focus on the SNR at the frequency of interest, $\Sigma$, as the noise is multiplicative and depends on the frequency index.
While the transmitted power per user is constant, the power associated to the maximum frequency changes with the number of transmitters $K$. By inspection it can be shown that the amplitude in the absence of noise is
\begin{equation}
    A_{\Sigma}=
    B_c^K
    \left(\frac{1}{2}\right)^{K-1}
    \left(\sqrt{\frac{2}{N}}\right)^{K-1}
    \label{eq:Power_sum}
\end{equation}
The first term corresponds to the $B_c$ gain provided by each transmitter; the second term arises due to the product of cosines (in a product of cosines, the component associated to the sum of frequencies concentrates half the power); the last term is associated to the normalization required to compute the inverse DCT. Equation \eqref{eq:Power_sum}
evinces that the modulation requires $B_c\geq\sqrt{2N}$ for a non-vanishing maximum frequency. 
The amplitude of both maximum tones in Fig. \ref{fig:logos_demodulation} is the same because the constraint is satisfied with equality.

Regarding the noise, consider \eqref{eq:FM_AirComp_noise} in the frequency domain:
\begin{equation}
    \mathbf{d}=\text{DCT}\{\mathbf{r}\}=\mathbf{Qr}=\mathbf{QZp},
    \label{eq:signal_dct_vector}
\end{equation}
where lowercase and uppercase bold symbols correspond to vectors and matrices, respectively. We define $\mathbf{Q}$ as the DCT matrix, $\mathbf{Z}=diag(\mathbf{z})$ and $\mathbf{p}=\exp\left(\mathbf{x}_k/A_c+\mathbf{x}_l/A_c\right)$.
To study the noise distribution we express \eqref{eq:signal_dct_vector} as signal plus noise:
\begin{equation}
    \mathbf{d}=\mathbf{s}+\bm{\varepsilon}=\mathbf{Qp} + \mathbf{Q(Z-I)p}
    \label{eq:signal_dct_vector_additive}
\end{equation}
The second term reflects that the noise term is correlated with the transmitted signal. Each entry of the noise in \eqref{eq:signal_dct_vector_additive} corresponds to
\begin{align}
    \varepsilon[i]=\sum_{n=0}^{N-1}\mathbf{p}[n]\mathbf{q}_i[n]\left(z[n]-1\right),
    \label{eq:noise_distribution}
\end{align}
where $\mathbf{q}_i[n]$ is the $n$-th entry of the $i$-th DCT basis. Each of the terms in the sum corresponds to a weighted shifted log-normal distributed as
\begin{align}
    \log\mathcal{N}\left(\log(\mathbf{p}[n]\mathbf{q}_i[n]), \frac{\sigma^2}{A_c^2},-\mathbf{p}[n]\mathbf{q}_i[n]\right),
\end{align}
where the last term corresponds to the shift in the distribution. There is existing literature on the probability distribution of a sum of log-normal random variables. In \cite{Cobb12} the authors propose to use the Fenton-Wilkinson method to estimate the parameters for a single log-normal distribution that approximates the sum of log-normals; in \cite{Wu05} they use the short Gauss-Hermite approximation of the moment generating function, while \cite{Nag02} uses an extended convolution. To the best of our knowledge, \cite{Ras02} is the only work on weighted sums of log-normals. Nevertheless, all the previous results rely on numerical integration methods. To the extend of our knowledge, there is no closed-form expression for the distribution of \eqref{eq:noise_distribution}.

We have observed that, in practice, the central limit theorem applies in this case and we can approximate $\bm{\varepsilon}[i]\sim \mathcal{N}(\mu_{\varepsilon_{[i]}}, \sigma^2_{\varepsilon_{[i]}})$.
We find that the mean and variance at the frequency index $i=\Sigma$ are
\begin{align}
  \mu_{\varepsilon} = A_\Sigma(\mu_z-1)\qquad
  \sigma^2_{\varepsilon} = \sigma_z^2P_p
  \label{eq:noise_demod_var}
\end{align}
where $P_p$ is the power of noiseless signal in \eqref{eq:FM_AirComp_noise}, this is,
\begin{equation}
    P_p = \frac{1}{N}\sum_{n=0}^{N-1} \exp\left(\frac{2}{A_c}\left(x_k[n] + x_l[n]\right)\right)
    \label{eq:power_noiseless}
\end{equation}
The corresponding proofs are omitted due to space constraints.
Equation \eqref{eq:power_noiseless} can be easily generalized for $K$ users as
\begin{equation}
    P_p = \frac{1}{N}\sum_{n=0}^{N-1} \exp\left(\frac{2}{A_c}\sum_{k=1}^{K}x_k[n]\right)
\end{equation}
Notice that $P_p$ increases exponentially with $K$, which limits the applicability of Log-FSK to massive access scenarios.

Finally, the SNR at frequency $\Sigma$ can be expressed as
\begin{equation}
    \text{SNR}_\Sigma=\frac{A_\Sigma^2}{P_p\sigma_z^2 + A_\Sigma^2(\mu_z-1)^2}
    \label{eq:snr_destination}
\end{equation}
and written in terms of $\text{SNR}_R$ as
\begin{align}
    \frac{1}{\text{SNR}_\Sigma}=&
    \frac{P_p}{2N}\exp\left(\frac{P}{\text{SNR}_R}\right)
    \left(\exp\left(\frac{P}{\text{SNR}_R}\right)-1\right)\nonumber\\
    & + 
    \left(\exp\left(\frac{P}{2\text{SNR}_R}\right)-1\right)^2
    \label{eq:snr_destination_v2}
\end{align}
Notice that the SNR definition in \eqref{eq:snr_destination} differs from the conventional, where the denominator contains the signals uncorrelated with the signal of interest. Nonetheless, in light of \eqref{eq:snr_destination_v2} we observe the destination SNR always increases with the received SNR and, thus, with transmitted power as well, which is a desirable property in communications. However, the growth is not always linear. In high-SNR ($\text{SNR}_R$) regime, the SNR at destination in dB can be approximated as
\begin{align}
    \text{SNR}_\Sigma (dB) =
    10\log_{10}(2N) + \text{SNR}_R - P_p - P,
    \label{eq:high_SNR}
\end{align}
where all magnitudes are expressed in dB. At high SNR the signal and noise exhibit a linear relationship, although these are still correlated. 
With the proposed design, this is, $B_c=\sqrt{2N}$ and $\alpha=2+\delta$, 
$P$ is always below 0 dB for a wide range of values of $N$. Thus, it can be ignored in the expressions above.
At high-SNR we observe that there is a 3 dB gain when doubling the number of samples, which corresponds to the first term in \eqref{eq:high_SNR}.

\begin{figure}[t]
\centering
\includegraphics[width=\columnwidth]{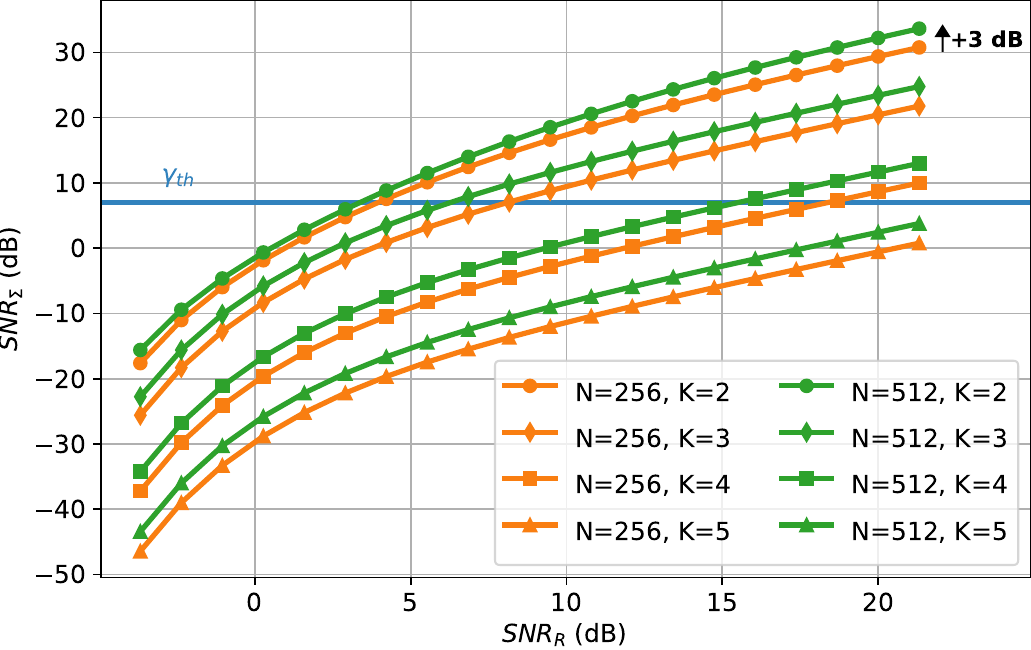}
\caption{SNR at destination ($\text{SNR}_\Sigma$) with respect to received SNR ($\text{SNR}_R$). The figure also shows the threshold at $\text{SNR}_\Sigma=7$ dB to achieve $P_e<\gamma_{th}=10^{-4}$.}
\label{fig:SNR_R_SNR_D}
\vspace{-0 pt}
\end{figure}

Fig. \ref{fig:SNR_R_SNR_D} shows the relationship between $\text{SNR}_R$ and $\text{SNR}_\Sigma$ for different number of samples $N$ and users $K$. Below the linear phase, Log-FSK exhibits a threshold effect under which $\text{SNR}_\Sigma$ drops rapidly. This is a natural behavior in any frequency modulated system, and a minimum $\text{SNR}_R$ is required to work above the threshold. 
The procedure to determine this threshold is the following: The performance of Log-FSK is determined by the correct detection of the peak at frequency $\Sigma$, which corresponds to a detection problem in $M$-ary FSK. We define $\gamma_{th}$ as the tolerable error probability and set the following constraint:
\begin{equation}
    P_e \approx (N-1)Q\left(\sqrt{\text{SNR}_{\Sigma}(N,K)}\right)\leq
    \gamma_{th},
    \label{eq: error_prob}
\end{equation}
where $P_e$ is the error probability in $M$-ary FSK and we explicitly show the dependence of $\text{SNR}_\Sigma$ with $N$ and $K$. Then, \eqref{eq: error_prob} is used to determine $\text{SNR}_\Sigma$ and \eqref{eq:snr_destination_v2} is used to find $\text{SNR}_R$ and the corresponding transmission power. Fig. \ref{fig:SNR_R_SNR_D} shows the threshold for $\gamma_{th}=10^{-4}$, which requires $\text{SNR}_\Sigma=7$ dB and is consistent with threshold values found in frequency modulations \cite{Car68}. While $P_p$ is independent of the transmitted power per user, the exponential relationship with respect to $K$ makes the high-SNR regime more challenging with increasing $K$. However, Log-FSK can still be used beyond $K=2$. For $K=3, 4$ and $5$ it requires around 7, 15 and 25 dB in transmission, respectively. 
Moreover, there are techniques to extend the threshold, such as phase locked loops \cite{Tho08} that can improve the performance of Log-FSK at low SNR.

\subsubsection{Mean Squared Error}
Since Log-FSK works as a detection system, the MSE is zero when $\Sigma$ is detected correctly; otherwise, the MSE is weighted by the error probability:
\begin{equation}
    \text{MSE}(N,K) = P_e\sum_{m=0}^{N-1}|m-\hat{\Sigma}|^2,
\end{equation}
where $\hat{\Sigma}$ is the incorrectly detected frequency. 





\subsection{Fading channels}
In the presence of flat fading channels, the effect needs to be compensated. Otherwise, the non-linear functions in the demodulation produces undersired effects. While frequency-based modulations do not require channel state information (CSI), Log-FSK requires CSI to transmit at the same power. The benefit with respect to standard AirComp is that it can be any power level, it does not need to be perfectly compensated.
In the case of frequency selective channels, phase compensation is needed as well.


\section{Numerical results}
\label{sec:sum}

\subsection{Number of simultaneous users}

While Log-FSK is not a modulation for massive access, the are scenarios in which network connectivity is low and the proposed system can be implemented.
For instance, in \cite{Guirado23} an AirComp scheme for intra-chip computations is implemented with only 3 transmitters. What is more, in some wireless sensor networks (WSN) the communication infrastructure is only possible between pairs of nodes. This is the case of linear and cyclical topologies: In \cite{Lee22} an AirComp scheme is developed for distributed vehicle platooning control. Since vehicles move along a highway, each one only communicates with its preceding and succeeding neighbors. A case of cyclical topologies is found in Low Earth Orbit (LEO) satellite constellations \cite{Leyva23}. In these networks satellites can communicate with 4 neighbors at most, this is, two from the same orbital plane and one for each adjacent plane.

\subsection{MSE comparison in averaging}
Computing the sum (or average) function is the standard example of AirComp systems, since it matches the additive nature of the communication channel. As in linear AirComp, Log-FSK also allows computing other functions via its nomographic representation \cite{GOldenbaum13}.

\begin{figure}[t]
\centering
\includegraphics[width=\columnwidth]{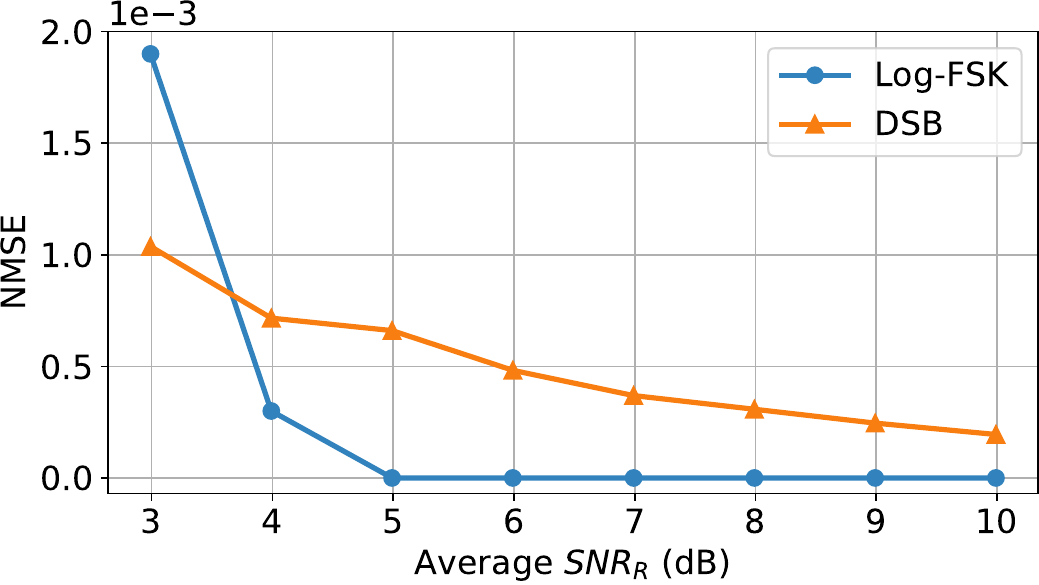}
\caption{NMSE with respect to average $SNR_R$ for Log-FSK and linear AirComp implemented with DSB.}
\label{fig:MSE_SNR_R}
\vspace{-0 pt}
\end{figure}

We consider a scenario with $K=2$ transmitters in an AWGN channel and we ignore the quantization noise. As a benchmark we choose the linear AirComp, implemented with the double sideband (DSB) modulation, this is, the information is encoded in the amplitude of the carrier. Since the $SNR_R$ in DSB depends on the transmitted information, we propose to evaluate both schemes in terms of average $\text{SNR}_R$, this is, at the same total transmitted power.
Fig. \ref{fig:MSE_SNR_R} displays the normalized MSE (NMSE) at different average $\text{SNR}_R$ regimes. The measurements $m_k$ come from a uniform distribution in $[0,N/2]$ and the NMSE is averaged over $10^4$ realizations. As expected, Log-FSK achieves zero MSE when working above the threshold SNR. Below the threshold DSB is recommended over Log-FSK, since in the former the noise affects linearly the demodulated information.

\section{Conclusions}
In this paper we propose Log-FSK, a novel frequency modulation for AirComp. The non-linear waveform design allows to recover a frequency component that corresponds to the sum of the individual transmitted frequencies.
We provide a demodulation procedure that relies on the DCT and extracting the maximum frequency.
We provide the theoretical performance of Log-FSK in terms of error probability and MSE, and evaluate the proposed scheme with respect to linear AirComp implemented with DSB. As expected, Log-FSK outperforms the benchmark with zero error when working above the threshold SNR. Finally, we propose use cases where the network connectivity meets the sparse multiple access that the system requires. Also, Log-FSK could facilitate physical layer network coding, where only two users transmit simultaneously.

\bibliographystyle{IEEEbib}
\bibliography{refs}

\end{document}